\documentclass[conference]{IEEEtran}
\IEEEoverridecommandlockouts
\usepackage{cite}
\usepackage{amsmath,amssymb,amsfonts}
\usepackage{graphicx}
\usepackage{textcomp}
\usepackage{xcolor}
\usepackage{tikz}
\usepackage{pgfplots}
\def\BibTeX{{\rm B\kern-.05em{\sc i\kern-.025em b}\kern-.08em
    T\kern-.1667em\lower.7ex\hbox{E}\kern-.125emX}}
    
    \usepackage{algorithm}
\usepackage{algpseudocode}

\let\vec\mathbf

\begin{document}

\title{A Novel Antenna Placement Algorithm for Compressive Sensing MIMO Radar\\
\thanks{This work was supported by the German Research Foundation (DFG, German Research Foundation) under Grant SFB 1483 – Project-ID 442419336.}
}

\author{\IEEEauthorblockN{Bastian Eisele\textsuperscript{$\ast$}, Ali Bereyhi\textsuperscript{$\ast$}, Ingrid Ullmann\textsuperscript{$\dagger$}, Ralf Müller\textsuperscript{$\ast$}}
\IEEEauthorblockA{\textsuperscript{$\ast$}Institute for Digital Communications (IDC), Friedrich-Alexander Universität Erlangen-Nürnberg (FAU) \\
\textsuperscript{$\dagger$}Institute of Microwaves and Photonics (LHFT), FAU\\
bastian.eisele@fau.de, ali.bereyhi@fau.de, ingrid.ullmann@fau.de, ralf.r.mueller@fau.de}
}

\maketitle

\begin{abstract}
	In colocated compressive sensing MIMO radar, the measurement matrix is specified by antenna placement. To guarantee an acceptable recovery performance, this measurement matrix should satisfy certain properties, e.g., a small \textit{coherence}. Prior work in the literature often employs randomized placement algorithms which optimize the prior distribution of antenna locations. The performance of these algorithms is suboptimal, as they can be easily enhanced via expurgation. In this paper, we suggest an iterative antenna placement algorithm which determines the antenna locations deterministically. The proposed algorithm locates jointly the antenna elements on the transmit and receive arrays, such that the coherence of the resulting measurement matrix is minimized. Numerical simulations demonstrate that the proposed algorithm outperforms significantly the benchmark, even after expurgation.
\end{abstract}


\section{Introduction}

Multiple-input-multiple-output (MIMO) radar employs multiple transmit and receive antennas for target detection \cite{li2008mimo}. Using multiple transmit antennas, it transmits independent (orthogonal) waveforms that can be separated at the receiver side. By doing so, the target scene is probed by several signals. MIMO radar systems are roughly divided into the following two categories: Distributed MIMO radar systems which use separated transmit and receive array antennas \cite{mimod}, and colocated MIMO radar systems which invoke dense arrays to perform signal transmission and reception by array antennas located physically together \cite{mimoc}.\par In MIMO radar systems, the location of antenna elements at the transmit and receive arrays specifies the measurement matrix by which the target object is sampled. To meet the desired performance in the system, this matrix often requires to satisfy certain properties. Such a requirement raises the problem of \textit{antenna placement} in MIMO radar systems in which the locations of the antenna elements are optimized with respect to a signal processing metric. \par This paper investigates the problem of antenna placement in a colocated MIMO radar system and proposes a robust antenna placement algorithm which guarantees efficient direction-of-arrival (DoA) estimation for sparse targets.
\subsection{Antenna Placement in Colocated MIMO Radar}
 To sample a signal at Nyquist-rate, a colocated MIMO radar systems should employ linear antenna arrays with sufficient aperture width \cite{wide_arrays}. To additionally avoid ambiguities, the distance between two neighboring antenna elements should not exceed half the wavelength \cite{mimoc}. This means that by using classical techniques, a fine resolution is achieved with rather large antenna arrays, and hence can impose high implementation costs. \par
 For sparse target scenes, one can exploit the sparsity of the target and employ compressive sensing (CS) to achieve a desired resolution with fewer antennas \cite{rossi}. In this case, the antenna elements on the transmit and receive arrays are placed irregularly and sparsely on the Nyquist aperture \cite{rossi}. Doing so, the radar system collects an underdetermined set of observations from which the DoA can be estimated via a sparse recovery algorithm \cite{cs_radar_book}. \par 
  The problem of DoA estimation with sparse irregular antenna arrays describes a standard underdetermined linear system whose measurement matrix is specified by the  positions of the antenna elements. From CS, we know that the choice of measurement matrix directly impacts the performance of the sparse recovery algorithm \cite{rauhut_book}. The measurement matrix hence should fulfill certain properties, i.e. a small coherence or restricted isometry property. This leads to the conclusion that the quality of the DoA estimation in a CS-based MIMO radar system depends on the antenna positioning \cite{rossi}. \par Several lines of work address the problem of measurement matrix construction in CS-based MIMO radar systems i.e. \cite{rossi} and \cite{b1,A_7,A_21,A_23,A_27}. In \cite{A_7,A_21,A_23} the measurement matrix is optimized by waveform design with respect to its coherence. Minimizing the measurement matrix coherence can also be done by antenna placement as it is shown in \cite{A_27} for distributed and in \cite{b1} for the colocated MIMO radar.
   In \cite{A_21} the authors introduce a power allocation and a waveform design scheme for both, colocated and widely separated CS-MIMO radar. The proposed schemes aim to minimize the difference between the Gram matrix of the CS measurement matrix and an identity matrix. A similar problem has been discussed in \cite{A_23} for colocated MIMO radar systems. Here, the optimization criteria is the coherence of the measurement matrix. For waveform design it is demonstrated that the coherence only depends on the covariance matrix of the transmitted waveforms. Consequently, the covariance matrix is optimized by methods of convex optimization and then transformed into realistic waveforms. In \cite{rossi}, the coherence of the measurement matrix is minimized by a random placement of the antenna elements from a proper probability distribution. By optimizing the prior distribution of antenna positions, measurement matrices with even smaller coherence are constructed in \cite{b1}. \par The study in this paper is mostly related to the recent work in \cite{b1} in which the authors aim to place a fixed number of antenna elements on predefined grids, such that the coherence of the resulting measurement matrix is minimized. To this end, the authors invoke the idea of random antenna placement via a proper prior distribution, and convert the original combinatorial problem into an alternative form in which an optimum prior distribution for antenna positions is to be derived, such that the \textit{expected} coherence of the measuring matrix is to be minimized. The alternative problem is then relaxed into a convex optimization which can be tractably solved. The antenna positions are then generated at random from the optimized distribution. 
     \subsection{Contributions and Organization}
      Using the idea of expurgation in random coding, one can easily improve the proposed algorithm in \cite{b1} by generating multiple random samples of antenna placements and choosing the one which leads to the minimal coherence. This modification results in a minor computational load\footnote{This is not of any concern.} and a considerable performance boost. Motivated by this observation, this paper proposes a novel antenna placement algorithm for CS-based radar systems. The proposed algorithm invokes a similar core-idea as in \cite{b1} and determines for each grid point on the transmit and receive array a selection probability. It however deviates from the randomized approach of \cite{b1} and updates the selection probabilities in an iterative fashion until they describe a deterministic antenna placement. Our numerical investigations show that the proposed algorithm outperforms the randomized approach of \cite{b1}, even after expurgation.   \par
The remaining of this manuscript is organized as follows. In Section \ref{section_CS}, we briefly go through the principles of CS. We then present the system model in Section \ref{section_SM}. In Section \ref{main}, we introduce the concrete problem setting, give a recap of the method from \cite{b1} and present the proposed algorithm. The performance of the proposed algorithm is then investigated through several numerical experiments in Section \ref{performance_chapter}. Finally, we conclude the paper in Section \ref{con}. 
\subsection{Notation}
We represent scalars, vectors and matrices as non-bold, bold lower-case, and bold upper-case letters, respectively. The notations $\vec{Q}^{\mathsf{T}}$ and $\vec{Q}^{\mathsf{H}}$ denote the transposed and transposed conjugate of matrix $\vec{Q}$, respectively. By $\vec{q}^{*}$, we refer to the element-wise conjugate of $\vec{q}$. Furthermore, $\vec{a}\odot \vec{b}$ and $\vec{a}\otimes \vec{b}$ denote Hadamard and Kronecker product of $\vec{a}$ and $\vec{b}$, respectively. Moreover, $[N]$ is used to abbreviate the set of integers $\{1,\ldots,N\}$ and for sets $A,B$, we use $A\setminus B$ to denote the set of elements contained in $A$ but not in $B$.
Finally, we write $\Vert \vec{v} \Vert_{p}$ to denote the $p$-norm of vector $\vec{v}$.
\section{Backgrounds on Compressive Sensing}
\label{section_CS}
CS deals with recovering a large sparse signal $\vec{x} \in \mathbb{C}^{n}$ from an underdetermined set of observations $\vec{z} \in \mathbb{C}^{m}$ ($m \ll n$) which are collected through a noisy linear system, i.e.  \cite{don}
\begin{equation}
\label{eq_1}
\vec{z} = \vec{A} \vec{x} + \vec{n}
\end{equation}
for some measurement matrix $\vec{A}\in \mathbb{C}^{m \times n}$ and additive noise $\vec{n}\in \mathbb{C}^{m}$. Without any prior information on $\vec{x}$, the task of recovering $\vec{x}$ from $\vec{z}$ is generally impossible, even in the noise-free case, i.e., $\vec{n}=\vec{0}$, due to the underdetermined nature of the problem. Using the sparsity of $\vec{x}$ as a prior information, CS can address this problem efficiently, if a certain set of constraints are satisfied on $\vec{A}$ \cite{candes}. We will write $\Vert \vec{v} \Vert_{0}$ to denote the sparsity of a vector $\vec{v}$, as its number of nonzero entries can be expressed by $\Vert \vec{v} \Vert_{p}$ for $p\rightarrow0$, see \cite{rauhut_book}. \par CS roughly deals with two major design tasks: (1) designing a reconstruction algorithm and (2) design of an efficient measurement matrix $\vec{A}$. In this paper, we deal with the second task: we intend to find an irregular antenna arrangement which leads to an efficient measurement matrix in the corresponding CS problem. \par 
There exist multiple measures to assess the quality of a measurement matrix, e.g., coherence, restricted isometry property and null spare property \cite{rauhut_book}. The design criteria in this paper is the so-called coherence\footnote{This is a typical choice of metric, due to its computational tractability.} which for a given matrix $\vec{A}$ is defined as
\begin{equation}
\label{coherence}
\mu(\vec{A}) = \max_{i\neq j} \frac{\left\vert\vec{a}_{i}^{\mathsf{H}}\vec{a}_{j} \right\vert}{\Vert\vec{a}_{i}\Vert\;\Vert\vec{a}_{j}\Vert},
\end{equation}
with $\vec{a}_{i}$ being the $i$-th column of $\vec{A}$. For a given sparse recovery algorithm, the coherence is desired to be small, in order to guarantee a good recovery performance. This can be interpreted as demanding the matrix columns to be as close as possible to a collection of orthogonal atoms. For further discussions on reconstruction guarantees in terms of the matrix coherence, see \cite{rauhut_book}.
\section{System Model}
\label{section_SM}
Consider a colocated MIMO radar system with linear antenna arrays, where $M$ transmit and $N$ receive antennas are placed on the $y$-axis at positions
\begin{subequations}
\begin{align}
	\vec{y}_{\mathrm{t}} &= [y_{t,1},\ldots,y_{t,M}],\\
	\vec{y}_{\mathrm{r}} &= [y_{r,1},\ldots,y_{r,N}],
\end{align}
\end{subequations}
respectively. Here, the $y$-axis is normalized to the wavelength $\lambda$. Targets are modeled as point-scatterers in the array far-field and described by their azimuth angle\footnote{The azimuth angle is measured counter-clockwise from the positive $x$-axis.} $\theta$ and reflection coefficient $\beta$. The transmit and array steering vectors for a target in azimuth angle $\theta$ can be described as
\begin{subequations}
\begin{align}
\vec{a}_{u}(\vec{y}_{\mathrm{t}}) &= \left[ e^{j2\pi u y_{t,1}}, e^{j2\pi u y_{t,2}},\ldots, e^{j2\pi u y_{t,M}}  \right]^{\mathsf{T}},  \\
\vec{b}_{u}(\vec{y}_{\mathrm{r}}) &= \left[ e^{j2\pi u y_{r,1}}, e^{j2\pi u y_{r,2}}, \ldots, e^{j2\pi u y_{r,N}}  \right]^{\mathsf{T}},
\end{align}
\end{subequations}
respectively. Here, we define $u=\sin \theta$ and refer to it as the DoA parameter. \par Consider a single point-scatterer at the DoA parameter $u$ with the reflection coefficient $\beta$. Assuming that the transmitter sends unit-power and mutually-orthogonal signals of length $M$ in a given domain, e.g., time or frequency domain, the augmented received signal $\vec{r}_{u} \in \mathbf{C}^{MN}$ can be derived as a sufficient statistic from the received signal as \cite{b1} 
\begin{equation}
\label{eq_2}
\vec{r}_{u} = \beta\:\vec{b}_{u}(\vec{y}_{\mathrm{r}}) \otimes \vec{a}_{u}(\vec{y}_{\mathrm{t}})+ \vec{n},
\end{equation}
where $\mathbf{n}$ is additive white Gaussian noise (AWGN) with zero mean and variance $\sigma^2$. Consequently, for a target with multiple point-scatterers, the receive signal can be written as the superposition of the signals received from each point-scatterer. \par We consider the scanning of a sparse object, as is commonly the case in practice. This means that we assume only $K$ point-scatterer on the predefined grid $\theta_{1},...,\theta_{G}$, with $G\gg K$, have non-zero reflection coefficients. As a result, the measurement model can be represented as
\begin{equation}
\label{system_model_eq}
\vec{r} = \vec{\Psi} \boldsymbol{\beta} + \vec{n}, 
\end{equation}
with $\boldsymbol{\beta}=[\beta_{1},...,\beta_{G}]^{\mathsf{T}}$ being the $K$-sparse vector of reflection coefficients. In \eqref{system_model_eq}, each column of $\vec{\Psi} \in \mathbb{C}^{MN\times G}$ is defined by the steering vectors of its corresponding DoA, i.e.,
\begin{equation}
\label{col_of_A}
\boldsymbol{\psi}_{g} = \vec{b}_{u_{g}}(\vec{y}_{\mathrm{r}}) \otimes \vec{a}_{u_{g}}(\vec{y}_{\mathrm{t}}) 
\end{equation}
for $g \in [G]$ and $u_{g} = \sin(\theta_{g})$.
\par For the particular sensing matrix in \eqref{system_model_eq}, the cross correlation between two column vectors $\boldsymbol{\psi}_{g}$ and $\boldsymbol{\psi}_{g^{'}}$, with $g \neq g'$ is given by \cite{b1}
\begin{equation}
\label{29582jrfwefwfs}
	\left\vert\boldsymbol{\psi}_{g^{'}}^{\mathsf{H}} \boldsymbol{\psi}_{g}\right\vert = \left\vert \vec{b}_{u_{g^{'}}}^{\mathsf{H}}(\vec{y}_{\mathrm{r}}) \vec{b}_{u_{g}}(\vec{y}_{\mathrm{r}}) \right\vert \left\vert \vec{a}_{u_{g^{'}}}^{\mathsf{H}}(\vec{y}_{\mathrm{t}}) \vec{a}_{u_{g}}(\vec{y}_{\mathrm{t}}) \right\vert.
\end{equation}
As a result, the calculation of the coherence can be split up between the transmit and receive array. \par We assume that all point-scatterers of the target whose reflection coefficients are non-zero, are placed on the grid $\theta_{1},...,\theta_{G}$. The angular grid ${\theta_{1},\ldots,\theta_{G}}$ is set non-uniformly on $[-\pi/2 , \pi/2]$ such that the DoA parameters of the grid, i.e., $u_{g}=\sin \theta_{g}$, form a uniform grid on the interval $[-1+\frac{2}{G},1]$. This means that, the grid in the $u$-domain is given by
\begin{equation}
	\label{u_grid_defi}
	\mathcal{U}_{G} = \{u_{1},\ldots,u_{G}\}, 
\end{equation}
where $u_{g} = -1+2g/G,$ for $g \in [G]$.
\par Our ultimate goal is to find the places of the antenna elements at the transmit and receive arrays, such that the coherence of the resulting measurement matrix is minimized. To this end, we consider a dense grid of possible antenna positions at the transmit and receive array from which the best combination of positions is to be found. More precisely, we assume that the $M$ transmit and $N$ receive antennas are located on predefined $\frac{\lambda}{2}$-spaced grids with $\tilde{M}$ and $\tilde{N}$ grid points, respectively, i.e., the positions of the transmit and receive antennas are chosen respectively from the following vectors of possible positions
\begin{subequations}
	\begin{align}
		\vec{\tilde{y}}_{\mathrm{t}} &= \left[0,\frac{\lambda}{2},\ldots,\frac{\lambda}{2} (\tilde{M}-1)\right], \\
		\vec{\tilde{y}}_{\mathrm{r}} &= \left[0,\frac{\lambda}{2},\ldots,\frac{\lambda}{2} (\tilde{N}-1)\right].
	\end{align}
\end{subequations}
For sake of compactness, we refer to the steering vectors of the full grids, i.e., $M=\tilde{M}$ and $N=\tilde{N}$ as $\vec{\tilde{a}}_{g} := \vec{a}_{u_{g}}(\vec{\tilde{y}}_{\mathrm{t}})$ and  $\vec{\tilde{b}}_{g} := \vec{b}_{u_{g}}(\vec{\tilde{y}}_{\mathrm{r}})$ for $g\in [G]$.
\section{Antenna Placement}
\label{main}
The antenna placement task can be formulated by introducing the weighting vectors $\vec{w}_{\mathrm{t}}\in \{0,1\}^{\tilde{M}}$ and $\vec{w}_{\mathrm{r}}\in \{ 0,1\}^{\tilde{N}}$ which specify the subset of selected grid points. Here, we set $w_{\mathrm{t},i}=1$ to refer to a selected grid point and $w_{\mathrm{t},i}=0$ to refer to an unselected point. Using these weighting vectors, our final goal can be reformulated as finding the jointly optimal $\vec{w}_{\mathrm{t}}$ and $\vec{w}_{\mathrm{r}}$, with $M$ and $N$ non-zero entries respectively, whose resulting measurement matrix has minimal coherence. This objective can be mathematically represented as
\begin{subequations}
\label{basic_problem}
\begin{align}
\min_{\vec{w}_{\mathrm{r}},\vec{w}_{\mathrm{t}}} \quad & \mu(\vec{\Phi}) \\
\textrm{s.t.} \quad & \sum_{i=1}^{\tilde{M}} w_{\mathrm{t},i} = M, \quad \sum_{j=1}^{\tilde{N}} w_{\mathrm{r},j} = N \\
& w_{\mathrm{t},i} \in \{0,1\},\; \forall i \in [\tilde{M}] ,\\  
& w_{\mathrm{r},j} \in \{0,1\},\; \forall j \in [\tilde{N}].  
\end{align}
\end{subequations}
Here, $\vec{\Phi}\in \mathbb{C}^{MN\times G}$ represents the effective measurement matrix for the given weighting vectors $\vec{w}_{\mathrm{t}}$ and $\vec{w}_{\mathrm{r}}$ whose column $g$ for $g\in[G]$ is given by the \textit{non-zero}\footnote{Note that $\boldsymbol{\phi}_{g} \in \mathbb{C}^{\tilde{M} \tilde{N} \times 1}$ has only $MN$ non-zero entries.} entries of
\begin{equation}
\label{15_423423}
\boldsymbol{\phi}_{g} = (\vec{w}_{\mathrm{r}} \odot \vec{\tilde{b}}_{g}) \otimes (\vec{w}_{\mathrm{t}} \odot \vec{\tilde{a}}_{g}).
\end{equation}
\par Considering \eqref{15_423423}, it is straightforward to show that the columns of $\vec{\Phi}$ have the following properties
\begin{enumerate}
	\item The Euclidean norms of all columns are fixed.
	\item The cross-correlation between the two columns $u$ and $u'$ only depends on the absolute difference between the indices, i.e., $\vert u-u'\vert$; see \cite{b1} for the detailed derivations.
\end{enumerate}
Using these two properties, the objective in \eqref{basic_problem} reduces to
\begin{equation}
\left\vert\boldsymbol{\phi}_{g^{'}}^{\mathsf{H}} \boldsymbol{\phi}_{g}\right\vert = \left\vert\vec{w}_{\mathrm{r}}^{\mathsf{T}} \vec{b}_{g^{'}g} \right\vert \; \left\vert\vec{w}_{\mathrm{t}}^{\mathsf{T}} \vec{a}_{g^{'}g} \right\vert,
\end{equation}
see \cite[Eqs. (23)-(26)]{b1}. Here, $\vec{a}_{g^{'}g}$ and $\vec{b}_{g^{'}g}$ are defined as
\begin{subequations}
\begin{align}
\vec{a}_{g^{'}g} &= \vec{a}_{g^{'}}^{*} \odot \vec{a}_{g}, \\ \vec{b}_{g^{'}g} &= \vec{b}_{g^{'}}^{*} \odot \vec{b}_{g}. 
\end{align}
\end{subequations}
Consequently, we pick an arbitrary index $g^{'}\in [G] $ and the antenna placement task is reduced to solving the following optimization:
\begin{subequations}
\label{basic_problem_reformulation}
\begin{align}
\min_{\vec{w}_{\mathrm{r}},\vec{w}_{\mathrm{t}}} \quad & \max_{g\in [G],g\neq g^{'} } \left\vert\vec{w}_{\mathrm{r}}^{\mathsf{T}} \vec{b}_{g^{'}g} \right\vert  \left\vert\vec{w}_{\mathrm{t}}^{\mathsf{T}} \vec{a}_{g^{'}g} \right\vert \\
\textrm{s.t.} \quad & \sum_{i=1}^{\tilde{M}} w_{\mathrm{t},i} = M, \quad \sum_{j=1}^{\tilde{N}} w_{\mathrm{r},j} = N \\
& w_{\mathrm{t},i} \in \{0,1\},\; \forall i \in [\tilde{M}] ,\\  
& w_{\mathrm{r},j} \in \{0,1\},\; \forall j \in [\tilde{N}].     
\end{align}
\end{subequations}
This optimization reduces to the NP-hard problem of integer programming and hence cannot be solved in a polynomial time.
\subsection{Randomized Iterative Antenna Placement Algorithm}
\label{section_A}
To approximate the solution of \eqref{basic_problem_reformulation}, the authors in \cite{b1} suggest a convex relaxation by letting the weighting vectors take any continuous values between $0$ and $1$ \cite[Eq. (30)]{b1}.
These real-valued weights are then interpreted as probabilities of the grid points being occupied by antennas. The relaxed problem can now be observed as optimizing antenna placement probabilities, such that the expected coherence is minimized. In contrast to \eqref{basic_problem_reformulation}, the relaxed problem no longer deteminestically fixes the number of transmit and receive antennas and only restricts their expectations. Though the relaxed version is no longer constrained by a non-convex set, it is still a hard problem to solve, since its objective is non-convex. This latter issue can be overcome by using alternating optimization \cite{b1}. To this end, we define
\begin{subequations}
\begin{align}
	a_{g^{'}g} &:= \left\vert\vec{w}_{\mathrm{t}}^{\mathsf{T}} \vec{a}_{g^{'}g} \right\vert, \\ b_{g^{'}g} &:= \left\vert\vec{w}_{\mathrm{r}}^{\mathsf{T}} \vec{a}_{g^{'}g} \right\vert.
\end{align}
\end{subequations}
Starting with an initial $\vec{w}_{\mathrm{t}}$, the solution of the joint optimization problem is then approximated by alternatively solving the convex optimization problem
\begin{subequations}
\label{resulting_relaxed_problem_a}
\begin{align}
\min_{\vec{w}_{\mathrm{r}}} \quad & \max_{g\in[G],g\neq g^{'} }  a_{g^{'}g}  \left\vert\vec{w}_{\mathrm{r}}^{\mathsf{T}} \vec{b}_{g^{'}g} \right\vert \\
\textrm{s.t.} \quad & \sum_{j=1}^{\tilde{N}} w_{\mathrm{r},j} = N, \;
 0 \leq w_{\mathrm{r},j} \leq 1 ,\; \forall  j \in [\tilde{N}]   
\end{align}
\end{subequations}
in terms of $\vec{w}_{\mathrm{r}}$, and the convex optimization problem
\begin{subequations}
\label{resulting_relaxed_problem_b}
\begin{align}
\min_{\vec{w}_{\mathrm{t}}} \quad & \max_{g\in [G],g\neq g^{'} }  b_{g^{'}g}  \left\vert\vec{w}_{\mathrm{t}}^{\mathsf{T}} \vec{a}_{g^{'}g} \right\vert \\
\textrm{s.t.} \quad & \sum_{i=1}^{\tilde{M}} w_{\mathrm{t},i} = M,\;
 0 \leq w_{\mathrm{t},i} \leq 1 ,\; \forall  i \in [\tilde{M}]    
\end{align}
\end{subequations}
for the solution of \eqref{resulting_relaxed_problem_a} in terms of $\vec{w}_{\mathrm{t}}$. The solution of the relaxed problem is finally approximated by the converging $\vec{w}_{\mathrm{t}}$ and $\vec{w}_{\mathrm{r}}$. As shown in \cite{b1}, these problems can be formulated and solved efficiently as second-order cone programs (SOCP) \cite{boyd_book}. For instance we can write \eqref{resulting_relaxed_problem_a} as
\begin{subequations}
\begin{align}
\min_{\vec{w}_{\mathrm{r}},t} \quad & t \\
\textrm{s.t.} \quad & a_{g^{'}g}  \left\vert\vec{w}_{\mathrm{r}}^{\mathsf{T}} \vec{b}_{g^{'}g} \right\vert < t, \forall g\in [G],g\neq g^{'}  \\
& \vec{1}^{\mathsf{T}} \vec{w}_{\mathrm{r}} = N, \;   
 0 \leq w_{\mathrm{r},j} \leq 1 ,\; \forall  j \in [\tilde{N}].   
\end{align}
\end{subequations}
\par In \cite{b1}, $\vec{w}_{\mathrm{t}}$ is initialized randomly and \eqref{resulting_relaxed_problem_a} and \eqref{resulting_relaxed_problem_b} iterate till convergence. The resulting weighting vectors $\vec{w}_{\mathrm{t}}$, $\vec{w}_{\mathrm{r}}$ are interpreted as probabilities from which the actual antenna placements are drawn at random. This scheme is summarized in \cite{b1} in Algorithm 1. In the sequel, we refer to it as the randomized iterative antenna placement (RIAP) algorithm. 
\subsection{Deterministic Iterative Antenna Placement Algorithm}
As mentioned in the introductory part, the iterative algorithm in Section \ref{section_A} can be easily boosted by expurgation. To this end, one can generate multiple realization of the random antenna placements, once the probabilities are found, and then choose the realization which results in minimal coherence. This leads to a minor increase in complexity, while it can reduce the coherence significantly. This observation motivates us to design a deterministic algorithm. \par In a nutshell, the proposed algorithm determines the antenna positions as follows: it uses the convex relaxation in \cite[Eq. (30)]{b1} and finds real-valued weights in $[0,1]$, by alternately optimizing the transmit and the receive array via \eqref{resulting_relaxed_problem_a} and \eqref{resulting_relaxed_problem_b}. The key difference to the RIAP algorithm is however in the alternation strategy: After each round of updating the weights $\vec{w}_{\mathrm{r}}$ and $\vec{w}_{\mathrm{t}}$,  we introduce a crucial new step in which we find the smallest entries of the resulting weighting vectors, and eliminate their corresponding grid points from the grid of possible antenna positions, $\vec{\tilde{y}}_{\mathrm{t}}$ or $\vec{\tilde{y}}_{\mathrm{r}}$. The intuition behind this step is to get rid of the grid slots that are least likely to be taken by an antenna according to their probabilities. Doing so, we gradually reduce the size of $\vec{\tilde{y}}_{\mathrm{t}}$ and $\vec{\tilde{y}}_{\mathrm{r}}$ until we are only left with $M$ positions for the transmit array and $N$ positions for the receive array. In contrast to RIAP this approach finds the antenna positions deterministically. As we show later on, the proposed approach outperforms the RIAP algorithm even after expurgation. \par We now present the algorithm formally. To this end, let us introduce the sets $\mathcal{J}_{\mathrm{t}} \subset [\tilde{M}]$ and $\mathcal{J}_{\mathrm{r}} \subset [\tilde{N}]$ which represent the sets of eliminated grid points at the transmit and receive arrays, respectively. Using this notation, we can define the marginal optimization for the evaluation of the weights at the receive grid points as
\begin{subequations}
\label{my_a}
\begin{align}
\min_{\vec{w}_{\mathrm{r}}} \quad & \max_{g\in [G],g\neq g^{'} }  a_{g^{'}g}  \left\vert\vec{w}_{\mathrm{r}}^{\mathsf{T}} \vec{b}_{g^{'}g} \right\vert \\
\textrm{s.t.} \quad & \sum_{j=1}^{\tilde{N}} w_{\mathrm{r},j} = N, \\
& 0 \leq w_{\mathrm{r},j} \leq 1 ,\; \forall  j \in [\tilde{N}] \setminus \mathcal{J}_{\mathrm{r}}     \\
& w_{\mathrm{r},j} = 0 ,\; \forall j \in \mathcal{J}_{\mathrm{r}}.   
\end{align}
\end{subequations}
Similarly, the marginal optimization problem which calculates the weights at transmit grid is modified as 
\begin{subequations}
\label{my_b}
\begin{align}
\min_{\vec{w}_{\mathrm{t}}} \quad & \max_{g\in [G],g\neq g^{'} }  b_{g^{'}g}  \left\vert\vec{w}_{\mathrm{t}}^{\mathsf{T}} \vec{a}_{g^{'}g} \right\vert \\
\textrm{s.t.} \quad & \sum_{i=1}^{\tilde{M}} w_{\mathrm{t},i} = M, \\
& 0 \leq w_{\mathrm{t},i} \leq 1 ,\; \forall  i \in [\tilde{M}] \setminus \mathcal{J}_{\mathrm{t}}     \\
& w_{\mathrm{t},i} = 0 ,\; \forall i \in \mathcal{J}_{\mathrm{t}}.   
\end{align}
\end{subequations}
The above optimization problems can be efficiently imposed as SOCP; see the discussions in \cite{b1}. For instance, \eqref{my_a} can be reformulated as 
\begin{subequations}
	\begin{align}
		\min_{\vec{w}_{\mathrm{r}},t} \quad & t \\
		\textrm{s.t.} \quad & a_{g^{'}g}  \left\vert\vec{w}_{\mathrm{r}}^{\mathsf{T}} \vec{b}_{g^{'}g} \right\vert < t, \forall g\in [G],g\neq g^{'}  \\
		& \vec{1}^{\mathsf{T}} \vec{w}_{\mathrm{r}} = N,    \\
		& 0 \leq w_{\mathrm{r},j} \leq 1 ,\; \forall  j \in [\tilde{N}] \setminus \mathcal{J}_{\mathrm{r}}     \\
		& w_{\mathrm{r},j} = 0 ,\; \forall j \in \mathcal{J}_{\mathrm{r}}.     
	\end{align}
\end{subequations}
\par In general, one can eliminate multiple grid points in each alternation. This leads to a clear trade-off: By eliminating multiple grid points, we reduce the complexity on one hand, and degrade the performance on the other hand. We hence introduce the parameter $p$ which controls the elimination of antenna positions. For a given $p$, the algorithm adds indices to $\mathcal{J}_{\mathrm{t}}$ and $\mathcal{J}_{\mathrm{r}}$ as long as the entries of the remaining indices of the transmit (or receive) weighting vector sum up to at least $M-p$ (or $N-p$). Essentially this parameter controls how fast grid points are ruled out from the total set of possible positions for antennas. \par The proposed scheme is summarized in Algorithm \ref{alg_my}. In the sequel, we refer to Algorithm \ref{alg_my} as the deterministic iterative antenna placement (DIAP) algorithm.
\begin{algorithm}[b]
	\caption{Deterministic Iterative Antenna Placement}\label{alg_my}
	\begin{algorithmic}[1]
		\State \textbf{Input:} $\vec{a}_{g^{'}g},\vec{b}_{g^{'}g} \; \forall g \in [G],M,N,p$
		\State \textbf{Initialize:} $\vec{w}_{\mathrm{t}}$ $\left(\Vert\vec{w}_{\mathrm{t}}\Vert_{1} = M\right)$, $\mathcal{J}_{\mathrm{t}}=[\:]$, $\mathcal{J}_{\mathrm{r}}=[\:]$
		\While{$\vert [\tilde{M}]\vert -\vert\mathcal{J}_{\mathrm{t}}\vert>M$ or $\vert [\tilde{N}]\vert - \vert\mathcal{J}_{\mathrm{r}}\vert>N$} 
		\State $a_{g^{'}g} = \vert\vec{w}_{\mathrm{t}}^{\mathsf{T}} \vec{a}_{g^{'}g} \vert$ $\forall$ $g \in [G]$, $g\neq g_{'}$
		\State Solve \eqref{my_a} for $\vec{w}_{r}$
		\While{$\Vert\vec{w}_{\mathrm{r}}\Vert_{1}>N-p$ and $\Vert\vec{w}_{\mathrm{r}}\Vert_{0}>N$}
		\State $\mathcal{J}_{\mathrm{r}}=\mathcal{J}_{\mathrm{r}} \cup \{j\}$, with $w_{\mathrm{r},j} = \min(\vec{w}_{\mathrm{r}}), j\not\in \mathcal{J}_{\mathrm{r}}$
		\State Set $w_{r,j} = 0$
		\EndWhile
		\State $b_{g^{'}g} = \vert\vec{w}_{\mathrm{r}}^{\mathsf{T}} \vec{b}_{g^{'}g} \vert$ $\forall$ $g \in [G]$, $g\neq g_{'}$
		\State Solve \eqref{my_b} for $\vec{w}_{\mathrm{t}}$
		\While{$\Vert\vec{w}_{\mathrm{t}}\Vert_{1}>M-p$ and $\Vert\vec{w}_{\mathrm{t}}\Vert_{0}>M$}
		\State $\mathcal{J}_{\mathrm{t}}=\mathcal{J}_{\mathrm{t}} \cup \{i\}$, with $w_{\mathrm{t},i} = \min(\vec{w}_{\mathrm{t}}), i\not\in \mathcal{J}_{\mathrm{t}}$
		\State Set $w_{\mathrm{t},i} = 0$
		\EndWhile
		\EndWhile
		\State \textbf{Output:} $\vec{w}_{\mathrm{t}},\vec{w}_{\mathrm{r}}$
	\end{algorithmic}
\end{algorithm}
\section{Performance Analysis}
\label{performance_chapter}
In this section, we analyze the performance of the DIAP algorithm and compare it to the RIAP method in \cite{b1}. To this end, we set $G=200$ and grid the transmit and receive arrays with $\tilde{M}=100$, $\tilde{N}=100$ positions, respectively. In contrast to the benchmark which considers a continuous initialization for $\vec{w}_{\mathrm{t}}$, we initialize the transmit weights with $w_{\mathrm{t},i}\in\{\ 0,1 \}$ and  $\Vert\vec{w}_{\mathrm{t}}\Vert_{1} = M$, as by experiment it appears to be favorable. \par
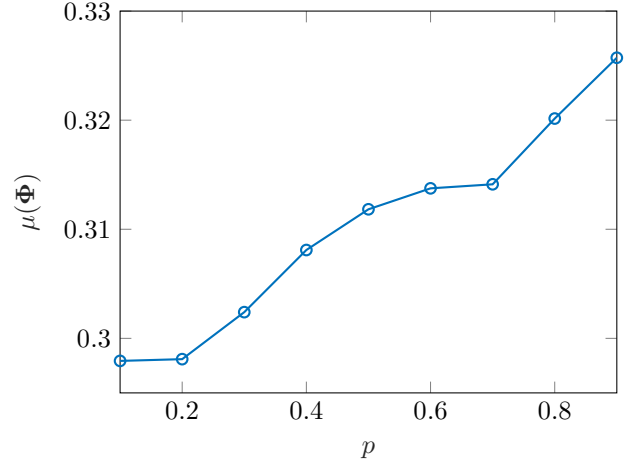
\begin{figure}[!t]
	\begin{center}
%
%
\definecolor{mycolor1}{rgb}{0.00000,0.44700,0.74100}%
\begin{tikzpicture}

\begin{axis}[%
width=2.6in,
		height=2in,
at={(0.758in,0.481in)},
scale only axis,
xmin=0.1,
xmax=0.9,
xlabel style={font=\color{white!15!black}},
xlabel={$p$},
ymin=0.295,
ymax=0.33,
ylabel style={font=\color{white!15!black}},
ylabel={$\mu(\vec{\Phi})$},
axis background/.style={fill=white},
legend style={legend cell align=left, align=left, draw=white!15!black}
]
\addplot [color=mycolor1,mark=o,style={thick}]
  table[row sep=crcr]{%
0.1	0.29793228101587\\
0.2	0.298092849388931\\
0.3	0.302401946368472\\
0.4	0.308097553602689\\
0.5	0.311829901448624\\
0.6	0.31375059998936\\
0.7	0.314121018484306\\
0.8	0.320148577221174\\
0.9	0.325722523425849\\
};

\end{axis}
\end{tikzpicture}%
	\end{center}
\caption{Average coherence of DIAP against $p$.}
\label{fig_1}
\end{figure}

  We speed up the DIAP algorithm by eliminating multiple grid points in each iteration. To this end, we conduct an experiment to determine a reasonable choice for the parameter $p$. As mentioned, a small choice of $p$ is expected to favor the performance, while setting $p$ to large values can reduce the run-time. We consider a setting with $M=7$ transmit and $N=7$ receive antenna elements and vary $p$ from $0.1$ to $0.9$ in steps of $0.1$. The resulting coherence is averaged over $100$ realizations and is plotted against $p$ in Fig. \ref{fig_1}.
We can observe that the performance of DIAP only slowly degrades with the growth of $p$. Noting that the complexity is not of a major concern in our setting, we set $p=0.33$. \par
\begin{figure}[!t]
	\begin{center}
%
%
\definecolor{mycolor1}{rgb}{0.00000,0.44700,0.74100}%
\definecolor{mycolor2}{rgb}{0.85000,0.32500,0.09800}%
\begin{tikzpicture}

\begin{axis}[%
width=2.6in,
		height=2in,
at={(0.758in,0.481in)},
scale only axis,
xmin=4,
xmax=14,
xlabel style={font=\color{white!15!black}},
xlabel={$M=N$},
ymin=0.1,
ymax=0.8,
ytick={0.1,0.2,0.3,0.4,0.5,0.6,0.7,0.8},
ylabel style={font=\color{white!15!black}},
ylabel={$\mu(\vec{\Phi})$},
axis background/.style={fill=white},
legend style={legend cell align=left, align=left, draw=white!15!black}
]
\addplot [color=mycolor1,mark=o,style={thick}]
  table[row sep=crcr]{%
4	0.736353901115955\\
5	0.626800933991646\\
6	0.541939617989016\\
7	0.46495507102633\\
8	0.418771988396655\\
9	0.368625397381203\\
10	0.337410967091981\\
11	0.281405026605868\\
12	0.270510318994219\\
13	0.244110689640557\\
14	0.223268428980466\\
};
\addlegendentry{RIAP}

\addplot [color=mycolor2,mark=o,style={thick}]
  table[row sep=crcr]{%
4	0.625594924575295\\
5	0.459166416469121\\
6	0.361997827682566\\
7	0.297567291013657\\
8	0.254558203544743\\
9	0.225445886662798\\
10	0.202255146843663\\
11	0.185229694228928\\
12	0.168378785003782\\
13	0.164875246367368\\
14	0.154762222232713\\
};
\addlegendentry{DIAP}

\end{axis}


\end{tikzpicture}%
	\end{center}
\caption{Comparing average coherence achieved via RIAP and DIAP.}
\label{fig_2}
\end{figure}
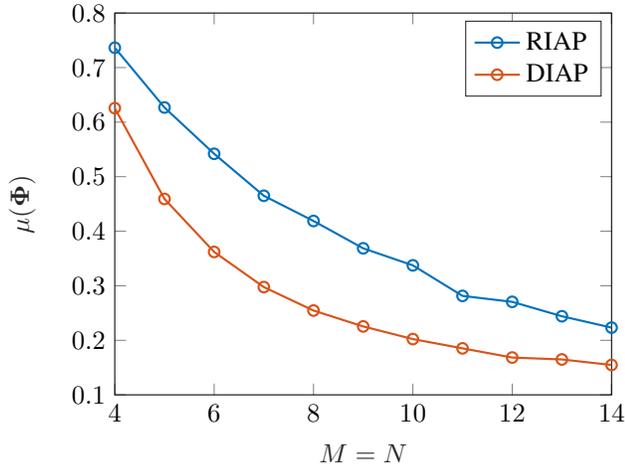

 To investigate the performance of DIAP, we vary the number of transmit and receive antennas from $4$ to $14$ while setting $M=N$. For each algorithm, we average the resulting coherence over $100$ runs for each choice of $M=N$. The result is shown in Fig. \ref{fig_2}, where we plot the coherence against the number of antenna elements.
As the figure shows, using DIAP algorithm, the coherence of the sensing matrix can be significantly reduced.\par
\begin{figure}[!t]
	\begin{center}
%
%
\definecolor{mycolor1}{rgb}{0.00000,0.44700,0.74100}%
\definecolor{mycolor2}{rgb}{0.85000,0.32500,0.09800}%
\definecolor{mycolor3}{rgb}{0.92900,0.69400,0.12500}%
\begin{tikzpicture}

\begin{axis}[%
width=2.6in,
		height=2in,
at={(0.758in,0.481in)},
scale only axis,
xmin=1,
xmax=20,
xlabel style={font=\color{white!15!black}},
xlabel={Iteration},
ymin=0,
ymax=100,
ylabel style={font=\color{white!15!black}},
ylabel={$\Vert\vec{w}_{\mathrm{t}}\Vert_{0},\Vert\vec{w}_{\mathrm{r}}\Vert_{0}$},
axis background/.style={fill=white},
legend style={legend cell align=left, align=left, draw=white!15!black}
]
\addplot [color=mycolor1,mark=o,style={thick}]
  table[row sep=crcr]{%
1	100\\
2	44.8\\
3	30.69\\
4	23.1\\
5	19.24\\
6	16.55\\
7	14.39\\
8	12.63\\
9	11.24\\
10	10.03\\
11	9.03\\
12	8.23\\
13	7.64\\
14	7.27\\
15	7.08\\
16	7.01\\
17	7\\
18	7\\
19	7\\
20	7\\
};
\addlegendentry{$\Vert\vec{w}_{\mathrm{t}}\Vert_{0}$}

\addplot [color=mycolor2,mark=o,style={thick}]
  table[row sep=crcr]{%
1	100\\
2	21.41\\
3	16.81\\
4	13.97\\
5	11.96\\
6	10.47\\
7	9.35\\
8	8.43\\
9	7.79\\
10	7.29\\
11	7.1\\
12	7.03\\
13	7.01\\
14	7\\
15	7\\
16	7\\
17	7\\
18	7\\
19	7\\
20	7\\
};
\addlegendentry{$\Vert\vec{w}_{\mathrm{r}}\Vert_{0}$}

\addplot [color=mycolor3, style={thick}]
  table[row sep=crcr]{%
1	7\\
2	7\\
3	7\\
4	7\\
5	7\\
6	7\\
7	7\\
8	7\\
9	7\\
10	7\\
11	7\\
12	7\\
13	7\\
14	7\\
15	7\\
16	7\\
17	7\\
18	7\\
19	7\\
20	7\\
};
\addlegendentry{$M=N$}

\end{axis}

\end{tikzpicture}%
	\end{center}
	\caption{Dimensionality of $\vec{w}_{\mathrm{t}}$ and $\vec{w}_{\mathrm{r}}$ against the iteration.} 
	\label{fig_3}
\end{figure}
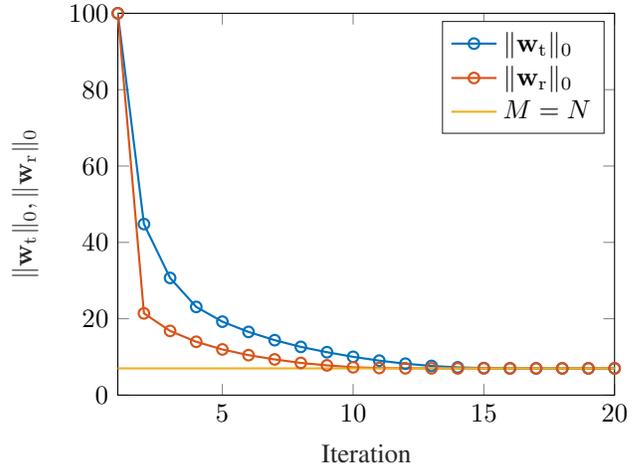

 To compare the complexity of both algorithms, we first plot the dimensinality of the optimization variables $\vec{w}_{\mathrm{t}}$ and $\vec{w}_{\mathrm{r}}$ depending on the iteration in Fig. \ref{fig_3}. Similar to the previous experiments, we set $M=N=7$ and average over $100$ realizations.
The figure indicates that the dimensionality quickly decreases in the first iterations and converges to the number of  transmit/receive antenna elements. Although the average number of iterations is higher than the one for the benchmark, the complexity of DIAP quickly reduces as it iterates, since the size of the optimization problems reduces in each alternation. In Table \ref{tabelle}, we further compare the overall run-time needed for both methods considering the same setting as in Fig. \ref{fig_3}. \par The results imply that the effective run-time is roughly doubled for $p=0.33$. However for DIAP the run-time can be significantly reduced by raising $p$ with only a moderate degradation in performance; see also Fig. \ref{fig_1}. We further demonstrate this behavior by setting $p$ to $p=1$ and $p=3$ and evaluate the run-time. Comparing to the benchmark, it is observed that DIAP is even faster while still achieving a significantly better coherence value.

\begin{table}
	\begin{center}
	\caption{Average coherence for RIAP and DIAP for different $p$}
	\label{tabelle}
		\scalebox{1.21}{
	\begin{tabular}{ c|c|c|c|c } 
 	   & RIAP & DIAP,   & DIAP, & DIAP,\\ 
 	   &     & $p=0.33$  & $p=1$ & $p=3$\\ 
  \hline
 $\mu(\vec{\Phi})$ & 0.47 & 0.30 & 0.33 & 0.37 \\ 
  \hline
 Runtime in [sec] & 2.23 & 4.56 & 3.08 & 1.89 \\ 
\end{tabular}
}
\end{center}
\end{table}




\section{Conclusion}
\label{con}
In this paper, we proposed an antenna placement scheme for colocated CS-MIMO radar. Unlike the earlier work in the literature which employs randomized antenna placement with optimized distribution, our approach determines the antenna positions deterministically. The results confirm our initial intuition based on expurgation indicating that by deterministic antenna placement the coherence of the measurement matrix can be significantly reduced. In its initial form, the proposed algorithm results in a slightly higher computational complexity. Nevertheless, the complexity-performance trade-off of the algorithm is easily controlled via parametrization. Interestingly, the proposed scheme always outperforms the benchmark, even when its complexity is set to be lower than the benchmark.


\bibliography{refer}
\bibliographystyle{IEEEtran}

\end{document}